\documentclass[%
 reprint,
 amsmath,amssymb,
 aps,
]{revtex4-2}

\newcommand{\Rayl}{\ensuremath{\mathrm{Ra}}}
\newcommand{\Tayl}{\ensuremath{\mathrm{Ta}}}
\newcommand{\Pran}{\ensuremath{\mathrm{Pr}}}
\newcommand{\xhat}{\ensuremath{\mathbf{\hat{x}}}}
\newcommand{\zhat}{\ensuremath{\mathbf{\hat{z}}}}

\newcommand{\Ty}{\ensuremath{T_y}}
\newcommand{\uvel}{\ensuremath{\mathbf{u}}}
\newcommand{\ubar}{\overline{\uvel}}
\newcommand{\uprime}{\uvel^\prime}
\newcommand{\aap}{Astronomy \& Astrophysics}

\usepackage{graphicx}
\usepackage{dcolumn}
\usepackage{bm}

\usepackage{upgreek}
\usepackage{cypriot}
\usepackage{linearb}
\usepackage{marvosym}
\usepackage[scr=boondox,  
            cal=esstix]   
           {mathalpha}

\begin{document}

\preprint{APS/123-QED}

\title{Ordering of timescales predicts applicability of quasi-linear theory in unstable flows}

\author{Curtis J. Saxton}
 \affiliation{Department of Physics, University of Warwick,
   Coventry  CV4~7AL, UK}
\author{Brad Marston}
\author{Jeffrey S. Oishi}
\altaffiliation[on leave from ]{Department of Physics \& Astronomy, Bates College, Lewiston, ME 04240}
\affiliation{
 Department of Physics, Box 1843, Brown University, Providence, RI, 02912-1843, USA
}%

\author{Steven~M. Tobias}%
 \email{S.M.Tobias@leeds.ac.uk}
\affiliation{%
 Department of Applied Mathematics, University of Leeds, Leeds LS2~9JT, UK
}%

\date{\today}

\begin{abstract}
We discuss the applicability of quasilinear-type approximations for a turbulent system with a large range of spatial and temporal scales. 
We consider a paradigm fluid system of rotating convection with a vertical and horizontal temperature gradients. 
In particular, the interaction of rotating with the horizontal temperature gradient drives a ``thermal wind'' shear flow
whose strength is controlled by a horizontal temperature gradient. 
Varying the parameters systematically alters the ordering of 
the shearing timescale, the convective timescale, and the correlation timescale. 
We demonstrate that quasilinear-type approximations work well when the shearing timescale or the correlation timescale is sufficiently short.
In all cases, the Generalised Quasilinear approximation (GQL) systematically outperforms the Quasilinear approximation (QL). We discuss the consequences for statistical theories of turbulence interacting with mean gradients.
We conclude with comments about the general applicability of these ideas across a wide variety of non-linear physical systems.

\end{abstract}

\maketitle

\section{Introduction}
Developing a description of nonlinear systems on a wide range of temporal and spatial scales is a significant  problem for nonlinear physics. 
One such canonical example is turbulence (in both fluids and plasmas) 
   where strong non-linearities lead to interactions on many scales. 
In its theoretically simplest manifestation of homogeneous and isotropic turbulence (HIT) 
   the goal is to develop a theory that can describe phenomena as diverse as 
   energy transfer via cascades, 
   development of structures, 
   and spatial and temporal intermittency.  
In this paper, we are concerned with geophysical and astrophysical flows that interact with instabilities, 
   often leading to anisotropic, inhomogeneous states and significant mean flows \citep{marston2023}.

On first view, this problem is less tractable than HIT; 
   however the interaction of fluctuating velocity with mean flows 
   may justify approximations that can lead to successful description of the flows.

The approximations usually involve the prioritisation of certain interactions over others 
   (and are formally exact when certain non-dimensional parameters are asymptotically small). 
These approximations are not only of interest because 
   they give information about which interactions are key in any given physical situation; 
   they may lead to the development of accurate closures for statistical theories describing fluids and plasmas.

The best-known example of such an approximation is the quasilinear (QL) approximation, 
   which dates back to Malkus~\cite{Malkus:1954dh}. 
He proposed that turbulence driven by an instability saturates 
   by modifying the mean state so that it is marginally stable to perturbations (on average). 
In this framework the primary interaction is that of the turbulent fluctuations  with themselves to modify the mean (back to marginality); 
   the interactions of fluctuations to produce other fluctuations (and a cascade leading to dissipation) are subdominant. 
Herring provided an early illustration in the context of convection \cite{herring1963investigation}.
However, this idea quite often does not work well, 
   though it can be shown to be formally correct in certain limits \citep{plumb77}.
In particular, the QL approximation works well when there is a strong separation of timescales,
   for example in the case of rapid rotation \citep{scottdritschel2012}, 
   strong stratification \citep{cmd2014}, 
   or for the cases of first-order smoothing in mean-field electrodynamics 
   \citep[see e.g.][]{MoffattDormy:2019,krauraed:1980,Tobias:2021}.

Recently, the quasilinear approximation has been extended 
   to consider more sets of non-linear, non-local interactions
   to obtain the generalized quasilinear (GQL) approximation. 
For details of this approximation see \citep{mct2016}. 
Briefly, the modes are spatially separated into low and high spectral wavenumbers 
   representing large and small horizontal spatial scales respectively; 
   the cut-off between low and high modes occurs at wavenumber $\Lambda$. 
In GQL, certain triad interactions are prioritised over others, 
   for example high/high $\rightarrow$ low interactions are kept 
   whereas high/high $\rightarrow$ high interactions are discarded. 
For $\Lambda = 0$ the system returns to the QL approximation described earlier 
   and as $\Lambda$ increases the system approaches full DNS. 
GQL has been shown to yield significant improvements over QL for even $\Lambda = 1$
\citep[see e.g.][]{child2016,hernandez2022a,oishi2023}.

In any driven, dissipative, turbulent system, 
   it is difficult to predict and control the important timescales in the flow 
   since they emerge as a result of the dynamics; 
   hence a separation of timescales that leads to effectiveness of quasilinear-type theories 
   (whether QL or GQL)
   can be difficult to enforce \textit{a priori}. 
In this paper, we propose a system where the ratio of timescales 
   can be modified by smoothly changing an input parameter: 
   the case of convection in the presence of a thermal wind. 
It is for this system we wish to investigate the utility of quasilinear theories.

We consider Boussinesq convection in a three-dimensional Cartesian domain rotating with angular velocity 
   $\boldsymbol{\Omega} = \Omega \boldsymbol{\hat{\Omega}}$
   at an angle $\uppi/2-\phi$ to the direction of gravity ($\zhat$), 
   as shown in Figure~\ref{fig:DNS}a. 
The fluid is subjected to both a vertical and horizontal temperature gradient
   and we non-dimensionalize lengths with the layer depth $d$, 
   time with the thermal diffusion time $\tau = d^2/\kappa$, 
   and temperature with a vertical temperature difference across the layer 
   $\beta d$ where $\beta = |\partial T/\partial z|$.
With this non-dimensionalization we write the non-dimensional basic state temperature as
\begin{equation}
    T_\textsc{b} = T_0 -\beta z + \Ty y,
    \label{eq:bs}
\end{equation}
and the basic state velocity in the $x$-direction owing to thermal wind balance \cite{hathaway1986,currie2014} is given by 
\begin{equation}
    U_\textsc{b} = -\frac{\Rayl\ \ \Ty}{\Tayl^{1/2} \sin \phi} \left( z - \frac{1}{2}\right),
\end{equation}
where the non-dimensional parameters are the Rayleigh number 
$\Rayl = g \alpha d^4 \beta/\kappa \nu$,
and Taylor number 
$\Tayl = 4 \Omega^2 d^4/\nu^2$.
In this paper we hold fixed $\Tayl = 10^5$ and Prandtl number 
   $\Pran = \nu/\kappa = 1$.

We solve for perturbations to the basic state, which satisfy  
\begin{align}
\frac{\mathrm{D} \uvel}{\mathrm{D} t} 
+\Tayl \Pran\ \boldsymbol{\hat{\Omega}} \times \uvel 
&= -\Pran \nabla p + \Rayl \Pran\ \theta \zhat + \Pran\ \nabla^2 \uvel \label{eq:mom}\\
\nabla \cdot \uvel &= 0 \label{eq:divu}\\
\frac{\mathrm{D}}{\mathrm{D} t} \left( T_\textsc{b} + \theta\right) &= \nabla^2 \theta,
    \label{eq:temp}
\end{align}
where the total convective derivative is given by
$\mathrm{D}/\mathrm{D}t \equiv 
\left(\partial /\partial t + (U_\textsc{b} \xhat + \uvel) \cdot \nabla\right) $.
The boundary conditions at $z=0,1$ are impenetrable, stress-free (on the velocity perturbations to the thermal wind), and isothermal, whilst the system is periodic in the horizontal directions.

\section{Direct Numerical Simulation and establishment of timescale regimes}

We initially perform fully nonlinear DNS of equations~(\ref{eq:mom}-\ref{eq:temp}) using the fully open source  {\tt Dedalus} pseudo-spectral code \citep{burns2019} using Fourier bases in the horizontal ($x$ and $y$) directions, and a Chebyshev representation in $z$. The horizontal dimensions are set as $L_x=5$, $L_y=5$. 
We perform simulations for three different choices of thermal wind
   ($\Ty = 0$, $-0.5$ and $-2$) 
   and two different choices of supercriticality  
   ($\Rayl = 4\times10^4$, and $2\times10^5$). 
We set the angle $\phi=\uppi/4$ so the box is considered to be at midlatitude. 
Resolutions are $(n_x,n_y,n_z)$ = $(64,64,64)$ for $\Rayl = 4\times10^4$;
$(64,64,128)$ for $\Rayl = 2\times10^5$, $\Ty=0$;
and $(128,128,64)$ for $\Rayl = 2\times10^5$, $\Ty=-2$.

\begin{figure}
    \centering
    \includegraphics[width=\textwidth]{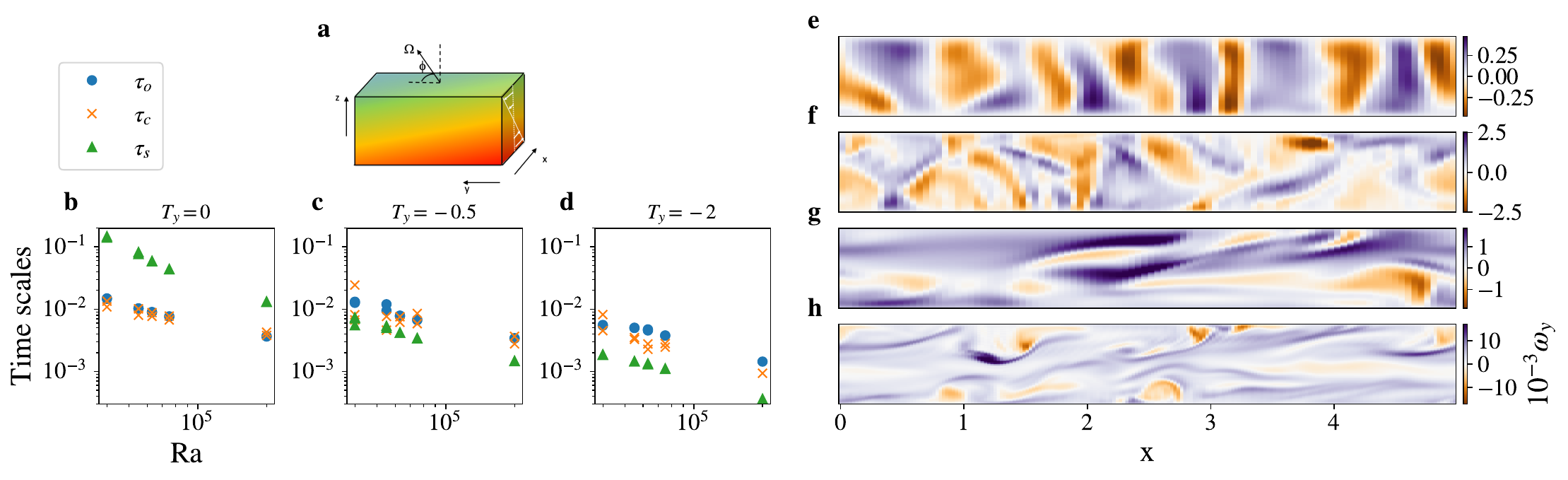}
    \caption{Results of DNS. a) sketch of domain, illustrating thermal wind and background temperature gradient;
    b--d) Overturning, correlation, and shear time scales $\tau_\mathrm{o}$, $\tau_\mathrm{c}$, $\tau_\mathrm{s}$
    as a function of $\Rayl$ for $\Ty = 0, -0.5, -2$ (left to right). e--f) $y$ component of vorticity for $\Ty = 0$: 
    $\Rayl = 4\times 10^4$ (e), $\Rayl = 2\times 10^5$ (f); 
    g--h) $y$ component of vorticity for $\Ty = -2$: $\Rayl = 4\times 10^4$ (g), $\Rayl = 2\times 10^5$ (h)}
    \label{fig:DNS}
\end{figure}

Figure~\ref{fig:DNS}e--h shows snapshots of the horizontal vorticity 
   in a plane for four parameter choices. 
For each case the temporal behaviour is chaotic 
   in the statistically steady state. 
Unsurprisingly, as the vertical thermal driving (as measured by $\Rayl$) 
   is increased the level of turbulence increases. 
Moreover, as the latitudinal thermal gradient is increased in magnitude, 
   the thermal wind increases (as shown by equation~\ref{eq:bs}) 
   and the solutions become more sheared. 
Thus horizontal vorticity increases in magnitude
   from figure~\ref{fig:DNS} e to g and from f to h.

Our hypothesis is that the efficacy of quasilinear-type approximations will vary as the ordering of timescales in the turbulence changes. 
A simple argument is to consider the nonlinear interactions in the momentum equation for the fluctuating turbulent velocity. 
For a velocity with mean flow $\overline{\uvel}$ and fluctuation $\uvel^\prime$ these interactions are given by a shear term 
   $\uprime \cdot \nabla \ubar$ 
   and an eddy/eddy $\Rightarrow$ eddy nonlinearity (EENL) term 
   $\uprime \cdot \nabla \uprime - \overline{\uprime \cdot \nabla \uprime}$ respectively. 

The scales of local spatial and temporal derivatives
   are characterized by correlation length $\ell_\mathrm{c}$ and time $\tau_\mathrm{c}$.
We define the convective turnover time $\tau_\mathrm{o}\approx \ell_\mathrm{c}/u_\mathrm{rms}$
   and a shear time  $\tau_\mathrm{s} \approx (\mathrm{d}\ubar/\mathrm{d}z)^{-1}$.
The ratio of the EENL term to the shearing term is given by $S = \tau_\mathrm{s}/\tau_\mathrm{o}$.
If $S$ is small then the EENL term is negligible compared with the shear term. 
Another interesting ratio of timescales is that which compares the importance the EENL term 
   to the local acceleration $\partial \uprime/\partial t$ 
   for which the timescale is the correlation time. 
This ratio is measured by the Kubo number $K=\tau_\mathrm{c}/\tau_\mathrm{o}$.
If $K$ is small then the turbulence rapidly decorrelates and so the EENL term may be discarded.

Figure~\ref{fig:DNS}b--d shows how these timescales change as a function of $\Rayl$ and $\Ty$.
In the case of no thermal wind (regular convection),
   the shear time $\tau_\mathrm{s}$ is long compared with both
   $\tau_\mathrm{o}=\ell_\mathrm{c}/u_\mathrm{rms}$
   and $\tau_\mathrm{c}$ (as calculated by \cite{kapyla2006}) 
   which are of the same order as each other. 
Hence $S$ is large and $K\sim {\cal O}(1)$ here. 
For the case with strong thermal wind $\Ty=-2$ 
   the situation has changed dramatically $\tau_\mathrm{s} \ll \tau_\mathrm{o}$
   and $\tau_\mathrm{c} < \tau_\mathrm{o}$.
Hence, for this case, $S \ll 1 $ is and $K<1$ here.
Our prediction, therefore, is that QL-type theories will work better for the case with a strong thermal wind.

\section{QL/GQL Results}
In order to assess the effectiveness of the quasilinear approximations, 
   we apply GQL approximations
   at different wavenumber threshold cutoffs $\Lambda$; 
   i.e. $k_x,k_y \le \Lambda$ are considered low modes 
   whilst all others are considered high modes. 
We evolve these models for all the parameter sets considered in the previous section. 
Here we consider $\Lambda=0$ 
   (corresponding to QL evolution in both horizontal directions) 
   and GQL evolution in the horizontal $\Lambda=1,5,10$.

\begin{figure}
    \centering
    \includegraphics[width=\textwidth]{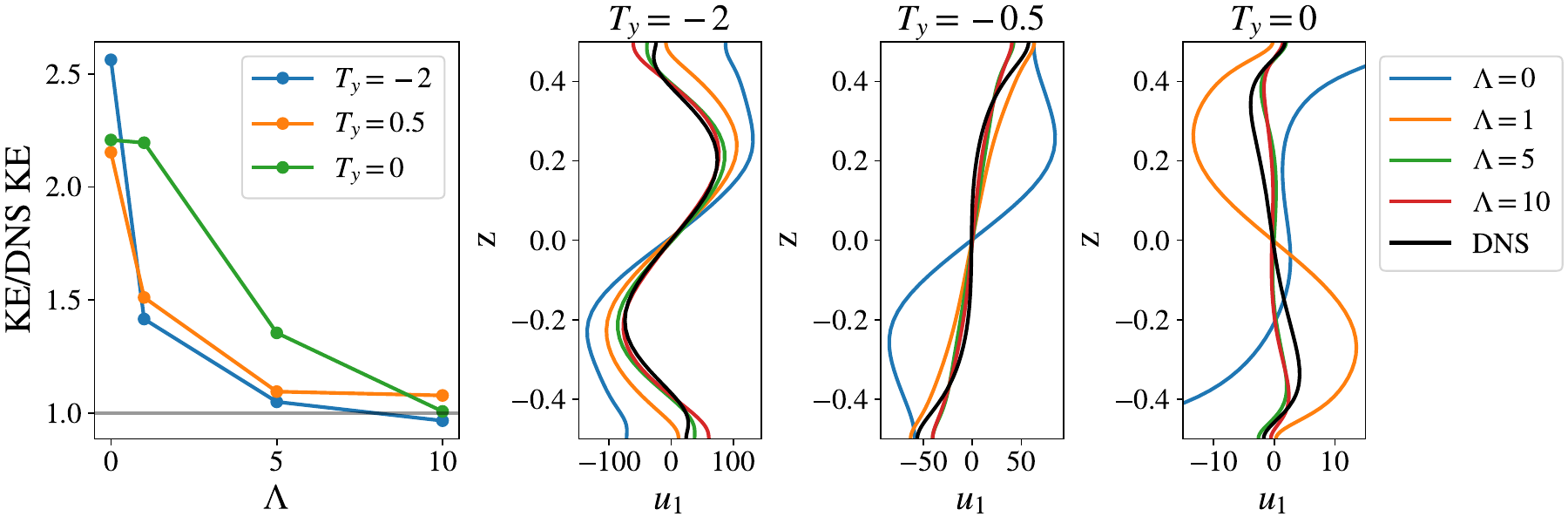}
    \caption{Results of QL/GQL simulations. Left: Time-averaged kinetic energy in saturation normalized by DNS kinentic energy as a function of GQL $\Lambda$ for $\Ty= -2, -0.5, 0$. Right: $z$ profiles of time and $x-y$ averaged $u$ velocity perturbations.}
    \label{fig:QL_GQL_results}
\end{figure}

Results are shown in Figure~\ref{fig:QL_GQL_results}. 
We calculate the total kinetic energy as a function of $\Rayl$, $\Ty$, and cutoff threshold $\Lambda$,
   and compare with DNS.  
Figure~\ref{fig:QL_GQL_results}a
   shows the average kinetic energy in the saturated state as a fraction of that obtained for DNS, for each $\Lambda$. 
The first thing to note is that all quasilinear approximations (for all the parameters) overestimate the kinetic energy. 
This is to be expected as the lack of the EENL term usually leads to underestimated dissipation, 
   owing to the missing local cascades.
For straight convection (no thermal wind; $\Ty=0$), 
   all the GQL approximations perform poorly until $\Lambda=10$. 
As $\Lambda$ increases, the approximation does improve, demonstrating that GQL performs better than QL ($\Lambda=0$). 
This agrees with previous studies that demonstrate how GQL can lead to improved performance over QL 
   by including energy transfer between high modes via non-local (in spectral space) interactions. 
This eddy-scattering effect is important.

The fact that GQL does not produce reasonable results until $\Lambda=10$ 
   shows that this system is not very amenable to quasilinear-type approximations 
   --- even if eddy-scattering off the mean flow is included --- 
   and that the EENL plays an important role. 
This is in agreement with our arguments in the previous section, 
   based on the ratio of timescales.

As $|\Ty|$ is increased, 
   the quasilinear approximations begin to perform better. 
In all cases, though, 
   the QL approximation ($\Lambda=0$) significantly overestimates the kinetic energies. 
However, when the shear time is small compared with the overturning time, 
   GQL (even with $\Lambda=1$) provides a good approximation to the full system 
   --- at least in representing global quantities such as the kinetic energy.

Figure~\ref{fig:QL_GQL_results}b--d shows how well the various levels of approximation 
   reproduce the mean flows profiles 
   (where the average is taken horizontally). 
Again for straight convection,
   the means are poorly represented by the quasilinear theories, 
   but as the separation of timescales becomes more pronounced 
   and the EENL term becomes subdominant then GQL, 
   even with $\Lambda=1$, 
   provides an accurate representation 
   of the saturated mean flow structures.

\begin{figure}
    \centering
    \includegraphics[width=\textwidth]{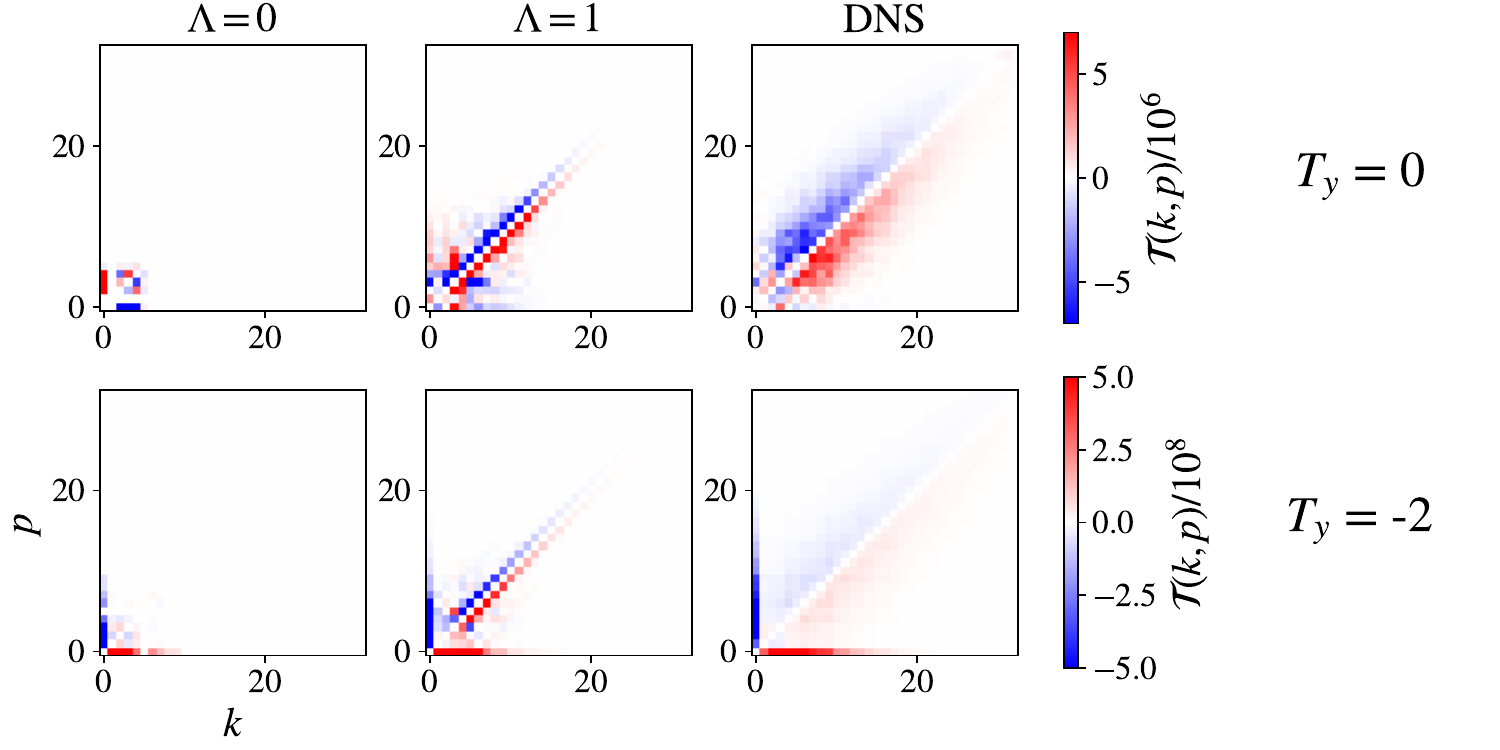}
    \caption{Spectral transfer function $\mathcal{T}(k, p)$ between two wave numbers $(k, p)$.
    Upper row shows $\Ty = 0$;
    lower row shows $\Ty = -2$. }
    \label{fig:transfer}
\end{figure}

The quasilinear approximation and its generalisations are examples of approximation via constrained triad decimation in pairs \citep{kraichnan1985}. The removal of certain interactions prevents certain routes for energy transfer. 
In order to calculate the preferred paths for transfer
   under the various degrees of approximation,
   we calculate energy transfer functions for the kinetic energy using methods outlined in
\cite{verma2004,alexakis2005,alexakis2007,lesur2011,favier2014,currie2020b}).
Figure~\ref{fig:transfer} shows these transfer functions for two different choices of $\Ty=0, -2$ 
   ($\Rayl$ is fixed at $2 \times 10^5$) 
   and two different choices of $\Lambda=0, 1$ and DNS. 
In all cases the QL version cuts off most of the channels for interaction 
   and so no cascades are available for energy transfer. 
Utilising GQL provides channels for non-local interactions and energy is scattered among modes.

The transfer function figure clearly shows that when $|\Ty|$ is small, 
   almost all transfers occur in triad interactions between fluctuating modes and the EENL term is important. 
The local cascade is therefore dominant for turbulent flows with no means. 
Note that in the case $\Ty=0$,
   QL saturates via interaction of turbulent fluctuations with the mean flow
   (upper left panel of Figure~\ref{fig:transfer}), 
   even though this is not the dominant mechanism in the DNS (upper right panel). 
However, when $\Ty=-2$  nearly all of the energy is transferred via the mean flow;  
   energy is transferred between turbulent modes in the cascade because of the interaction with the mean. 
It is this interaction that is more readily captured via quasilinear descriptions. 
A similar improvement in the accuracy of QL was found
   for the case of Langmuir turbulence which also breaks horizontal isotropy \cite{Skitka:2020co}.  

\section{Conclusions}

In this letter
   we investigated how the effectiveness of the quasilinear (QL)
   and generalised quasilinear (GQL) approximations in describing turbulent behaviour 
   depends on the ordering of timescales in the flow. 
We hypothesise that if the advective time of the turbulence
   is long compared with either a shearing time or a correlation time,
   then these quasilinear-type approximations will perform better.

We test these ideas by examining a rotating convective system in the presence of a thermal wind. 
This system is carefully chosen as the ordering of the timescales is naturally selected by the amplitude of the latitudinal temperature gradient 
   (for other parameters fixed). 
Using full Direct Numerical Simulation,
   as well as QL and GQL simulations we confirm our hypothesis. 
We also show that GQL systematically improves on QL dynamical descriptions.

What we have shown here is that,
   by constructing a fluid system with a tunable ordering of timescales,
   nonlinear interactions on a wide range of spatial and temporal scales
   can be understood and modeled via restricted modal equation sets. 
Although we have shown this for a turbulent fluid dynamic system, 
   the conclusion applies to a wide variety of quadratically nonlinear systems. 
As the restricted modal equation set consists of
   a reduction of interactions via triad decimation in pairs, 
   it conserves all global quadratic invariants in the limit of no driving and dissipation. 
We believe therefore that our approach is very general 
   and that progress can be made 
   on similar nonlinear problems with complicated spatial and temporal interactions.

\begin{acknowledgments}
We would like to acknowledge support of funding from
   the European Research Council (ERC)
   under the European Union's Horizon2020
   research and innovation programme
   (grant agreement no. D5S-DLV-786780).
We acknowledge the use of ARC supercomputer facilities
   at the University of Leeds.
This work has made use of NASA's Astrophysics Data System.
CJS thanks Anna Guseva for her months preserving the data presented here.
\end{acknowledgments}

\onecolumngrid

\twocolumngrid



\bibliography{refs_ann_rev}

\end{document}